\documentclass[superscriptaddress,twocolumn,prl]{revtex4}

\usepackage{amssymb}
\usepackage{amsmath}
\usepackage{graphicx}
\usepackage{epsfig}

\begin{document}

\title{Kinematical spin-fluctuation pairing in cuprates}

\author{\underline{N. M. Plakida}}
\email{plakida@theor.jinr.ru} \affiliation{Joint Institute for Nuclear Research,
 141980 Dubna, Russia}
\author{V.S. Oudovenko}
\affiliation{Rutgers University,  New Jersey 08854, USA }

\maketitle

We propose a microscopic theory of superconductivity for systems
with strong electron correlations such as cuprates in the
framework of  the extended Hubbard model where the intersite
Coulomb repulsion  and electron-phonon interaction are taken into
account. The Dyson equation for the normal and pair Green
functions for the Hubbard operators (HOs) is derived. Due to the
unconventional commutation relations for the  HOs, a specific
kinematical interaction of electrons with spin and charge
fluctuations with a large coupling constant of the order of the
kinetic energy of electrons $W$ emerges that results  in the
$d$-wave pairing with high-$T_c$. Superconductivity can be
suppressed only for a large intersite Coulomb repulsion $V
\gtrsim W$. Isotope effect  on $T_c$ caused by electron-phonon
interaction is weak at optimal doping and increases at low
doping. The kinematical interaction is absent in the spin-fermion
models and is lost in the slave-boson (-fermion) models treated
in the mean-field approximation.

\section{Introduction}

To explain unconventional  properties of cuprates one should take into account that
cuprates are  the Mott-Hubbard (more accurately, charge-transfer) doped insulators which
cannot be described within the conventional band theory (for a review see, e.g.,
~\cite{Plakida10}).  Under doping, a two-subband strongly-correlated metal emerges where
the Fermi-liquid model fails to describe electronic excitations.  The projected-type
(Hubbard) operators referring to the two subbands must be introduced. A new energy scale
of the order of the kinetic energy  of electrons $W$ arises  in the intraband hopping
induced by the kinematical interaction for the HOs which is much larger than the
antiferromagnetic exchange interaction $J$ induced by the interband hopping proposed by
Anderson~\cite{Anderson87}.  As shown in recent experiments~\cite{LeTacon11,LeTacon13},
short-range antiferromagnetic (AF) dynamical spin fluctuations survive in superconducting
state  even in the overdoped compounds. This justifies the spin-fluctuation mechanism of
superconductivity proposed earlier  within spin-fermion models (see, e.g.,
Refs.~\cite{Monthoux94b,Moriya00,Chubukov04,Abanov08}). We consider the spin-fluctuation
mechanism of pairing induced by the kinematical interaction where the coupling constant
is given by hopping parameters.

\section{Kinematical interaction in the Hubbard model}

To describe electronic systems with strong correlations the Hubbard model is commonly
used~\cite{Hubbard63}.   We consider the extended Hubbard model  on a square lattice
\begin{eqnarray}
H &= &\sum_{i \neq j, \sigma} \, t_{ij} \, a_{i\sigma}^{\dag} a_{j\sigma} + \frac{U}{2}
\,\sum_{i} N_{i \sigma} N_{i \bar\sigma} + H_{c, ep},
 \label{1}
\end{eqnarray}
where $a^{\dag}_{i\sigma}$ and $a_{i\sigma}$ are the Fermi creation and annihilation
operators for  electrons with spin $\sigma/2 \;(\sigma = \pm 1, \, \bar\sigma = -\sigma)
$ on the lattice site $i$, and $ \, N_{i} = \sum_{\sigma}\,N_{i\sigma} = \sum_{\sigma}\,
a^{\dag}_{i\sigma} a_{i\sigma} \, $ is  the number operator. $t_{ij}$ is the electron
hopping parameters (the nearest-neighbor hopping parameter  $t = 0.4$~eV is used  as the
energy unit). The on-site Coulomb interaction (CI) is $U$. The intersite CI $V_{ij}$ and
electron-phonon interaction (EPI) $g_{ij}$ are defined by the Hamiltonian:
\begin{eqnarray}
H_{c, ep} &= & \frac{1}{2}  \sum_{i\neq j}\,V_{ij} N_i N_j + \sum_{i, j}\,g_{i j} N_i\,
u_j,
 \label{1a}
\end{eqnarray}
where $u_j$ are  atomic displacements in  particular phonon modes. In the strong
correlation limit, $\,U \gg t\,$, the projected electron operators referring to the
single and double occupied subbands, the HOs, should be introduced~\cite{Hubbard65}:
 \[\,
  a^\dag_{i\sigma} =
a^\dag_{i\sigma}(1- N_{i \bar\sigma}) + a^\dag_{i\sigma} N_{i \bar\sigma} \equiv
X_{i}^{\sigma 0} + X_{i}^{2\bar\sigma}\,.
 \]
In terms of the HOs the model (\ref{1}) reads
\begin{eqnarray}
 H &= & \varepsilon_1\sum_{i,\sigma}X_{i}^{\sigma \sigma}
  + \varepsilon_2\sum_{i}X_{i}^{22} + \sum_{i\neq j,\sigma}\,
t_{ij}\,\bigl\{ X_{i}^{\sigma 0} \, X_{j}^{0\sigma}
\nonumber \\
& + &    X_{i}^{2 \sigma}X_{j}^{\sigma 2}
 + \sigma \,(X_{i}^{2\bar\sigma} X_{j}^{0 \sigma} +
  {\rm H.c.})\bigr\} + H_{c, ep},
 \label{2}
\end{eqnarray}
where $\varepsilon_1 = - \mu\,$ is the single-particle energy,
 $\varepsilon_2 =  U - 2\, \mu \,$ is the two-particle energy,
and $\mu \, $ is the chemical potential. The HO $X_{i}^{\alpha\beta} =
|i\alpha\rangle\langle i\beta|$ describes transition from the state $|i,\beta\rangle$ to
the state $|i,\alpha\rangle$ on the lattice site $i$ where $(\alpha, \beta ) $ refer to
four possible  states: an empty state $(\alpha, \beta =0) $, a singly occupied state
$(\alpha, \beta = \sigma)$, and a doubly occupied state $(\alpha, \beta = 2) $. The
number operator and the spin operators in terms of the HOs are defined as
\begin{eqnarray}
  N_i &=& \sum_{\sigma} X_{i}^{\sigma \sigma} + 2 X_{i}^{22},
\label{3a}\\
S_{i}^{\sigma} & = & X_{i}^{\sigma\bar\sigma} ,\quad
 S_{i}^{z} =  (\sigma/2) \,[ X_{i}^{\sigma \sigma}  -
  X_{i}^{\bar\sigma \bar\sigma}] .
\label{5}
\end{eqnarray}
The HOs obey the completeness relation $\,
  X_{i}^{00} +
 \sum_{\sigma} X_{i}^{\sigma\sigma}  + X_{i}^{22} = 1,
\,$ which rigorously preserves the constraint  that at any lattice site $i$ only one
quantum state $\alpha$ can be occupied. From the multiplication rule for the HOs $\,
X_{i}^{\alpha\beta} X_{i}^{\gamma\delta} = \delta_{\beta\gamma} X_{i}^{\alpha\delta}\,$
follows their commutation relations
\begin{equation}
\left[X_{i}^{\alpha\beta}, X_{j}^{\gamma\delta}\right]_{\pm}=
\delta_{ij}\left(\delta_{\beta\gamma}X_{i}^{\alpha\delta}\pm
\delta_{\delta\alpha}X_{i}^{\gamma\beta}\right)\, ,
 \label{3}
\end{equation}
with the upper  sign  for the Fermi-type operators (such as $X_{i}^{0\sigma}$) and the
lower sign for the Bose-type operators (such as  the number (\ref{3a}) or  spin (\ref{5})
operators).

The unconventional commutation relations (\ref{3}) for HOs result in the so-called {\it
kinematical} interaction introduced by Dyson  in a general theory of spin-wave
interactions ~\cite{Dyson56}. To demonstrate the role of the kinematical interaction in
the model (\ref{2}) let us consider an equation of motion for the HO  $\,
X\sb{i}\sp{\sigma 2} = a^{\dag}_{i\sigma}a_{i\sigma}a_{i\bar\sigma}\, $ :
\begin{eqnarray}
 i\frac{d}{d t}  X\sb{i}\sp{\sigma 2} &= &[X\sb{i}\sp{\sigma 2}, H] =
   (U - \mu )\, X_{i}^{\sigma 2} + [X\sb{i}\sp{\sigma 2}, H_{c, ep}]\,
\nonumber \\
  &+& \sum\sb{l,\sigma '}t\sb{il} \left(
    B\sb{i\sigma\sigma '}\sp{22} X\sb{l}\sp{\sigma ' 2} -
    \sigma \, B\sb{i\sigma\sigma '}\sp{21}
    X\sb{l}\sp{0\bar\sigma '} \right)
\nonumber \\
 &-& \sum\sb{l} t\sb{il}\, X\sb{i}\sp{02}  \left(
    X\sb{l}\sp{\sigma0} +  \sigma
    X\sb{l}\sp{2 \bar\sigma} \right),
 \label{4}
\end{eqnarray}
where the  Bose-type operators are introduced
\begin{eqnarray}
  B\sb{i\sigma\sigma'}\sp{22} & = & (X\sb{i}\sp{22} +
   X\sb{i}\sp{\sigma\sigma}) \, \delta\sb{\sigma'\sigma} +
   X\sb{i}\sp{\sigma\bar\sigma} \, \delta\sb{\sigma'\bar\sigma}
\label{4a} \\
  &  = &( N\sb{i}/2 +  \sigma\, S\sb{i}\sp{z}) \, \delta\sb{\sigma'\sigma} +
    S\sb{i}\sp{\sigma} \, \delta\sb{\sigma'\bar\sigma},
\nonumber \\
  B\sb{i\sigma\sigma'}\sp{21} & = & ( N\sb{i}/2 +
   \sigma S\sb{i}\sp{z}) \, \delta\sb{\sigma'\sigma} -
   S\sb{i}\sp{\sigma}\,  \delta\sb{\sigma'\bar\sigma},
  \label{4b}
\end{eqnarray}
We see that the hopping amplitudes depend on  number and spin operators caused by the
kinematical interaction.  In phenomenological spin-fermion models a dynamical coupling of
electrons with spin fluctuations is specified by fitting parameters (see,
e.g.,~\cite{Monthoux94b,Moriya00,Chubukov04,Abanov08}), while in Eq.~(\ref{4}) the
interaction is determined by the hopping energy $t_{ij}$ fixed by the electronic
dispersion.

\section{General formulation}

We consider superconducting pairing in the Hubbard model (\ref{2}) in the hole doping
region. In this case the  chemical potential $\mu$ is  situated in the two-hole upper
Hubbard subband and is determined by the equation for the average number of holes, $\, n
= 1 + \delta = \langle N_i \rangle \geq 1\,$.

To study the electronic spectrum and superconductivity in the model  we introduce the
two-time anticommutator Green function (GF)~\cite{Zubarev60} expressed in terms of the
four-component Nambu operators, $\, \hat X_{i\sigma}$ and $\, \hat
X_{i\sigma}^{\dagger}=(X_{i}^{2\sigma}\,\, X_{i}^{\bar\sigma 0}\,\, X_{i}^{\bar\sigma
2}\,\, X_{i}^{0\sigma}) \,$ for two subbands:
\begin{eqnarray}
 {\sf G}_{ij\sigma}(t-t') = \langle \!\langle \hat X_{i\sigma}(t) \mid
    \hat X_{j\sigma}^{\dagger}(t')\rangle \!\rangle.
 \label{k1}
\end{eqnarray}
To calculate the GF (\ref{k1})  we use the equation of motion method by differentiating
the  GF with respect to  time $t$ and $t'$. Using the projection operator
method~\cite{Plakida11} we derive the Dyson equation for the GF (\ref{k1})
~\cite{Plakida13}:
\begin{equation}
 {\sf G}\sb{\sigma}({\bf k}, \omega) =
  \left[\omega \tilde{\tau}\sb{0} - {\sf E}\sb{\sigma}({\bf k})
  -    {\sf  Q} {\sf \Sigma}_{\sigma}({\bf k}, \omega)
  \right] \sp{-1} {\sf Q},
\label{k2}
\end{equation}
where $\tilde{\tau}\sb{0}$ is the $4\times 4$ unit matrix and ${\sf Q} = \langle \{\hat
X\sb{i\sigma},\hat X\sb{i\sigma}\sp{\dagger}\}\rangle$.  The electron excitation spectrum
in the generalized mean-field approximation (GMFA) is determined by the time-independent
matrix of correlation functions:
\begin{equation}
  {\sf E}_{\sigma}({\bf k})=\frac{1}{N} \sum_{\bf j}
   {\rm e}^{i{\bf k (i-j)}}
\langle \{ [\hat X\sb{i\sigma}, H],
    \hat X\sb{j\sigma}\sp{\dagger} \} \rangle  {\sf Q}^{-1}.
\label{k2a}
\end{equation}
The self-energy operator  is given by the multiparticle GF,
\begin{equation}
 {\sf  Q} {\sf \Sigma}\sb{\sigma}({\bf k}, \omega) =
    \langle\!\langle {\hat Z}\sb{{\bf k}\sigma}\sp{(\rm ir)} \!\mid\!
     {\hat Z}\sb{{\bf k}\sigma}\sp{(\rm ir)\dagger} \rangle\!\rangle
      \sp{(\rm pp)}\sb{\omega}\;{\sf  Q}\sp{-1}.
\label{k3}
\end{equation}
The irreducible operators $\, \hat Z\sb{i\sigma}\sp{(\rm ir)}= [\hat X\sb{i\sigma}, H] -
\sum\sb{l}{\sf E}\sb{il\sigma} \hat X\sb{l\sigma}\,$ is determined by the equation  $\,
\langle \{Z\sb{i\sigma}\sp{(\rm ir)},\hat X\sb{j\sigma}\sp{\dagger} \} \rangle = 0 $.

To calculate the self-energy matrix (\ref{k3}) we use the self-consistent Born
approximation (SCBA) for the corresponding time-dependent multiparticle correlation
functions. Assuming an independent propagation of Fermi-type excitations $\, X_l^{\sigma'
2}, \,$ and Bose-type excitations $B\sb{i\sigma\sigma'}$  on different lattice sites  we
write the time-dependent multiparticle correlation functions as a product of fermionic
and bosonic correlation functions:
\begin{eqnarray}
\langle X_{m}^{2\sigma'} B\sb{j\sigma\sigma'}^\dag
 |B\sb{i\sigma\sigma'}(t)
X_l^{\sigma' 2}(t)\rangle|_{m \neq j, \, i \neq l}
\nonumber \\
 = \langle X_{m}^{2\sigma'} X_l^{\sigma' 2}(t)\rangle \langle
B\sb{j\sigma\sigma'}^\dag
 |B\sb{i\sigma\sigma'}(t) \rangle
 \label{k3a}
\end{eqnarray}
The time-dependent single-particle correlation functions are calculated self-consistently
using the corresponding GFs. This approximation results in a self-consistent system of
equations for the self-energy  (\ref{k3}) and the  GF (\ref{k2}).

The GFs for two subbands $1 (2)$ in the normal state in the imaginary frequency
representation can be written as
\begin{eqnarray}
\{G_{1(2)}({\bf k},  \omega_{n})\}^{-1}= i\omega_n  - {\varepsilon}_{1(2)}({\bf k}) -
\Sigma({\bf k},\omega_n) ,
 \label{k5}
\end{eqnarray}
where $ {\varepsilon}_{1(2)}({\bf k})$ are the quasiparticle energy  (\ref{k2a}) in GMFA.
The self-energy for  the two subbands  can be approximated by the same  function:
\begin{eqnarray} {\Sigma}({\bf k}, \omega_{n}) & = & -
\frac{T}{N}\sum_{\bf q}
 \sum_{m}\lambda^{(+)}({\bf q, k-q} \mid
\omega_{n}-\omega_{m})
 \nonumber \\
&\times &    [{G}_{1}({\bf q}, \omega_{m})+
  {G}_2({\bf q}, \omega_{m})]
\nonumber\\
 &\equiv & i\omega_{n}\,[1-Z_{\bf k}(\omega_n)]
  + X_{\bf k}(\omega_n).
 \label{k4}
\end{eqnarray}
In Fig.~1 we show the  doping dependence of  $Z({\bf q}) = Z_{\bf q}(0) $ which weakly
depends on $\delta$ in the underdoped case for $\delta \lesssim 0.15$ but sharply
decreases in the overdoped region.
\begin{figure}
\begin{center}
\includegraphics[width=0.5\linewidth]{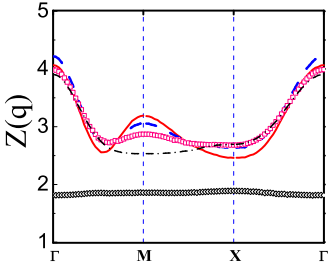}
\caption{(Color online) Doping dependence of the renormalization parameter $Z({\bf q}) $
along the symmetry directions $\Gamma(0, 0)\rightarrow M(\pi,\pi) \rightarrow X (\pi, 0)
\rightarrow \Gamma(0, 0)$ at $T \approx 140$~K for $\delta = 0.05$  (red solid line),
$\delta = 0.10$ (blue dashed line), $\delta = 0.15$ (pink squares), $\delta = 0.25$
(black dash-dotted line), and $\delta =  0.35$ (black diamonds)~\cite{Plakida13}.}
 \end{center}
\end{figure}

\section{Gap equation and $\bf T_c$}

Using the equation for the anomalous (pair) GF we derive equation for the superconducting
gap function~\cite{Plakida13}. In the linear approximation,  for the gap  in the two-hole
subband $\varphi_{\bf k}(\omega)= \sigma \varphi_{2, \sigma}({\bf k},\omega)$  we obtain
the equation
\begin{eqnarray}
\varphi({\bf k},\omega_n) =
  \frac{T_c}{N}\sum_{{\bf q}, m}\,\frac{[1 - b({\bf q})]^2\,
  \varphi({\bf q}, \omega_{m})}{[\omega_m Z_{\bf q}(\omega_m)]^2
  + [{\varepsilon}_{2}({\bf q}) + X_{\bf q}(\omega_m)]^2}
\nonumber \\
\times\{ J({\bf k-q}) - V({\bf k-q}) +\lambda^{(-)}({\bf q, k-q} \mid
\omega_{n}-\omega_{m})\},\quad
 \label{k5a}
 \end{eqnarray}
where $b({\bf q})$ takes into account the hybridization effects of the two Hubbard
subbands.  $J({\bf q})$  and  $V({\bf q})$ are the exchange interaction  and the
inertsite CI.  The frequency-dependent interaction is given by the function
\begin{eqnarray}
\lambda^{(\pm)}({\bf q },{\bf k } | \nu_n) = - |t({\bf q})|^{2} \, \chi\sb{sf}({\bf
k},\nu_n)
\nonumber \\
\mp [|V({\bf k})|^2 \,\chi_{cf}({\bf k}, \nu_n)
 + |g({\bf k })|^2 \, \chi_{ph}({\bf k}, \nu_n)],
\label{k5b}
 \end{eqnarray}
where $\,t({\bf q})$ is  the Fourier component of the hopping parameter $t_{ij}$. The
spectral density of spin $(sf)$, charge $(cf)$ fluctuations and phonons $(ph)$ are given
in terms of the dynamical susceptibility by the relations~\cite{Zubarev60}:
$\chi\sb{sf}({\bf q},\omega) =   -  \langle\!\langle {\bf S\sb{q} | S\sb{-q}}
\rangle\!\rangle\sb{\omega}\,$,  $
 \chi\sb{cf}({\bf q},\omega) =
 - \langle\!\langle \delta N\sb{\bf q} | \delta N\sb{-\bf q}
   \rangle\!\rangle\sb{\omega}$, and $
\chi\sb{ph}({\bf q},\omega) =
 - \langle\!\langle  u\sb{\bf q} | u\sb{-\bf q}
   \rangle\!\rangle\sb{\omega}$.
\par
For comparison of various contributions to the pairing we consider the gap equation close
to the Fermi energy, $\varphi({\bf k}) = \varphi({\bf k},\omega =0)$. In this case
instead of the dynamical susceptibility the static susceptibility  $\chi({\bf q}) ={\rm
Re} \, \chi({\bf q},\omega = 0) $ appears in the  gap equation:
\begin{eqnarray} &&\varphi({\bf k}) =
  \frac{1}{N}\sum_{{\bf q}} \,
  \frac{[1 - b({\bf q})]^2\,
  \varphi({\bf q})}{[Z({\bf q})]^2 \; 2\widetilde{\varepsilon}({\bf q})}
    \tanh\frac{\widetilde{\varepsilon}({\bf q})}{2T_c}
\nonumber \\
&& \times\big\{ J({\bf k-q})- V({\bf k-q})+  |V({\bf k -q})|^2 \chi\sb{cf}({\bf k -q})
 \nonumber \\
&&+ |g({\bf k -q})|^2 \, \chi_{ph}({\bf k-q})\,\theta(\omega_{0}-|
\widetilde{\varepsilon}({\bf q})|)
\nonumber \\
&& -  |t({\bf q})|\sp{2}\; \chi_{sf}({\bf k -q})\theta(\omega_{s} -|
\widetilde{\varepsilon}({\bf q})|) \big\}\,,
     \label{k5c}
\end{eqnarray}
where $\,  \widetilde{\varepsilon}({\bf q}) = \varepsilon_{2}({\bf q})/Z({\bf q}) $ is
the renormalized energy. Here $\omega_0 = 0.1t$ and $\omega_{s} = J = 0.4\, t$ are  the
cutoff energies for phonon and  spin-fluctuation excitations. Note, that in cuprates
$J({\bf q}) \lesssim V({\bf q})$ and therefore the AF exchange interaction $J({\bf q})$
cannot provide superconducting pairing proposed by Anderson~\cite{Anderson87}.

To solve Eq.~(\ref{k5c}) we should introduce models  for static susceptibilities.  The
spin susceptibility is determined by the function
\begin{eqnarray}
&&  \chi_{sf}({\bf k}) =\frac {\chi_{ Q}}{1+ \xi^2 [1+ \gamma({\bf k})]}  ,
 \label{k6}
\end{eqnarray}
where $\gamma({\bf k}) = (1/2)(\cos k_x + \cos k_y)$ and $\xi $ is the  AF correlation
length. The strength of the spin-fluctuation interaction is given by the susceptibility
at the AF wave vector ${\bf Q = (\pi,\pi)}$, $\chi_{ Q} = \chi_{sf}({\bf Q}) $,
 which is fixed by the normalization condition $\,
\langle {\bf S}_{i}^2\rangle
  = (3/4)(1- \delta)\,$:
\begin{equation}
\chi_{ Q} = \frac{3 (1- \delta)}{2\,\omega_{s} }
 \left\{ \frac{1}{N} \sum_{\bf q}
 \frac{1}{ 1+\xi^2[1+\gamma({\bf q})]} \right\}^{-1} .
 \label{k7}
 \end{equation}
For the EPI  we adopt a model  with strong forward scattering proposed in
Ref.~\cite{Zeyher96}. The static interaction in the model  is determined by function:
\begin{equation} g_{ep}({\bf k}) = |g({\bf k })|^2 \, \chi_{ph}({\bf
k}) = g_{ep}\,\frac{\xi_{ch}^2}{1 + \xi_{ch}^2 \,k^2} ,
 \label{k8}
\end{equation}
where the doping dependent parameter $\xi_{ch} = 1/ (2 \delta )$ determines the radius of
a ``correlation hole''.

\begin{figure}
\begin{center}
\includegraphics[width=0.5\linewidth]{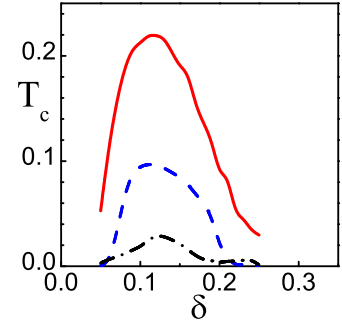}
\caption{(Color online) $T_c(\delta)$ in the WCA induced by all interactions (red solid
line) and only by the spin-fluctuation contribution $\widehat{\chi}\sb{sf}$(blue dashed
line) or only by the EPI $\, \widehat{V}\sb{ep}$ (black dash-dotted
line)~\cite{Plakida13}. }
 \end{center}
\end{figure}

\begin{figure}[h]
\begin{center}
\includegraphics[width=0.5\linewidth]{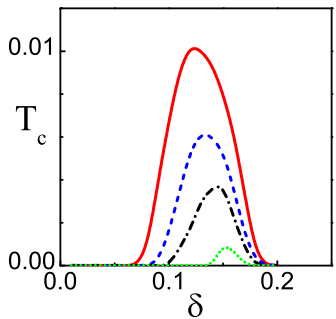}
\caption{(Color online) $T_c(\delta)$ for $V = 0.0$  (bold red line), $V = 0.5 $ (blue
dashed line), $V = 1.0 $ (black dash-dotted line), and $V = 2.0 $ (green dotted line) for
$U = 8 \,$~\cite{Plakida14}. }
\end{center}
 \end{figure}

To estimate various contributions in the gap equation (\ref{k5c}) we consider the
$d$-wave model gap, $\varphi({\bf k}) = (\Delta/2) \,(\cos k_x - \cos k_y)$. At first we
consider solution of the gap equation (\ref{k5c}) for $T_c$ in the weak coupling
approximation (WCA), $Z({\bf q})=1\,$. As shown in Fig.~2, the largest contribution comes
from the spin-fluctuation pairing induced by the kinematical interaction which is given
by the  averaged over the Fermi surface constant   $\, g_{sf} = \langle|t({\bf
q})|\sp{2}\; \chi_{sf}({\bf k -q})\rangle_{\rm FS} \approx 4 \,t \sim 2$~eV. Charge
fluctuations and EPI contributions appear to be small since only the angular momentum
$l=2$  of these interactions give contributions to the $d$-wave pairing~\cite{Plakida13}.
\par
In the strong coupling approximation (SCA) in Eq.~(\ref{k5a}) values of $T_c$ are reduced
by an order of magnitude as shown in Fig.~3 in comparison with the WCA in Fig.~2 due to
large values of renormalization parameter $Z({\bf q}))$ in Fig.~1. $T_c$ dependence on
the nearest neighbor intersite CI $V$ in Fig.~3 reveals that the $d$-wave pairing
survives as long as the Coulomb repulsion $V$ does not exceed the kinematical interaction
of the order of the kinetic energy, $ V \lesssim 4 \, t\,$,
\begin{figure}[h]
\begin{center}
\includegraphics[width=0.5\linewidth]{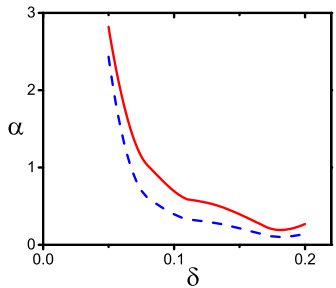}
 \caption{(Color online) Isotope exponent $\alpha(\delta)$ for $g_{ep} = 5\, t$ (red
solid line) and  $g_{ep} = 2.5\, t$  (blue dashed  line).}
 \end{center}
 \end{figure}

To study the isotope effect on $T_c$ within the gap equation (\ref{k5c}) we consider the
mass-dependent phonon frequency $\omega_{0} = \omega_{0}^{(0)}\sqrt{ {M_0}/(M_0 +\Delta
M)} = \omega_{0}^{(0)}(1 - \beta)$. We neglect the polaronic effect for the $d$-wave
electron-phonon coupling constant  (\ref{k8}) (for discussion see
Ref.~\cite{Plakida11a}). The result of numerical solution of the gap equation (\ref{k5c})
for the oxygen isotope exponent $(\beta = 1/16)$ $\, \alpha = - {d \,  \log \, T_{\rm
c}}/{ d \, \log \, M}\,$ is shown in Fig.~4 for two values of EPI $g_{ep} = 5\, t$ and
$g_{ep} = 2.5\, t\,$ in Eq.~(\ref{k8}). In accordance with an analytical estimation,
$\alpha  = (1/2)\,{g_{ep}}/{(g_{ep} + g_{sf} )}$, $\alpha$ increases for larger values of
$g_{ep}\,$. The doping dependence of the exponent agrees with experiments (see, e.g.,
\cite{Khasanov04}): it is quite small, $\, \alpha= 0.09 - 0.18\,$,  close to the optimal
doping  while drastically increases in the underdoped case at $\delta < 0.1$, $\, \alpha
= 0.38 - 0.68\,$  for $g_{ep} = 2.5\, t - 5\, t\,$, respectively. Similar results were
obtained in
Ref.~\cite{Shneyder09}) within the $t$-$J$ model with  EPI.\\

To summarize, we have shown that in the limit of strong correlations  a new coupling
parameter, of the order of the kinetic energy of electrons, appears in the two-subband
regime for the Hubbard model and brings about  $d$-wave superconductivity with
high-$T_c$. This kinematical interaction is lost in the spin-fermion models and the
slave-boson (-fermion) models treated in MFA.

\end{document}